\documentclass[sort&compress,5p]{elsarticle}
\usepackage{graphicx,natbib,amssymb}
\usepackage[pdftex,colorlinks,citecolor=blue,bookmarks]{hyperref}
\usepackage{bm}
\usepackage{amsmath}
\usepackage{color}

\journal{Physics Letter B}

\begin{document}
	\begin{frontmatter}
		\title{On the robustness of sub-shell closures: A high angular momentum analysis of the titanium isotopes.} 
		
		\author[UAM]{Tom\'as R. Rodr\'iguez}
		\ead{tomas.rodriguez@uam.es}
		\author[UAM]{J. Luis Egido\corref{cor1}}
		\ead{j.luis.egido@uam.es}
		\cortext[cor1]{Corresponding author}
		
		\address[UAM]{Departamento de Física Teórica y Centro de Investigación Avanzada en Física Fundamental, Universidad Autónoma de Madrid, E-28049 Madrid, Spain}

\begin{abstract}

The potential sub-shell closures  $N=32$ and $N=34$ are analyzed  at high spins in the titanium isotopes within the generalized beyond mean field theory considering  triaxial deformations and the angular frequency as generator coordinates together with the particle number and the angular momentum conservation.  A good description of bulk properties, high angular momenta spectra and transition probabilities is obtained. The outcome at high spin in these nuclei is consistent with the magic number character of $N=32$  but not of  $N=34$. 

\end{abstract}


\end{frontmatter}

Magic numbers are landmarks of the nuclear physics landscape. Their study opened the door to the nuclear shell model and to its interpretation as a manifestation of shell closures \cite{MJ.55}. Presently, with the new radioactive beam facilities, one is able to explore the regions far away from the stability line allowing to study the persistence or disappearance of the {\em classical} magic numbers and the emergence of new ones~\cite{PPNP_61_602_2008}. In the search of these new magic numbers one has been guided by the nuclear masses, by the excitation energy of the $2^{+}_1$ state, and/or by the E2 decay of this state to the ground state. Hence, a sudden drop in nucleon separation energies, a larger $2^{+}_1$ excitation energy and a smaller transition probability as compared to the neighbor isotones (or isotopes) are commonly considered as manifestations of the magic character of the nucleus. An indication of the robustness of a new shell closure is given by its extension through nearby isotopes (or isotones). 

In this way the possible shell closures at $N=32$ and $N=34$ (coming from the filling of the $1p_{3/2}$ and $1p_{1/2}$ spherical orbits, respectively) have been extensively studied in the past. The $N=32$ gap determined in calcium ($Z=20$)~\cite{PRC_31_2226_1985,Nature_498_346_2013} has been found to persist in argon ($Z=18$)~\cite{PRL_114_252501_2015}, potassium ($Z=19$)~\cite{PRL_114_202501_2015}, scandium ($Z=21$)~\cite{PRC_79_014313_2009}, titanium ($Z=22$)~\cite{PLB_546_55_2002, PRC_70_064303_2004, PRC_70_064304_2004} and chromium ($Z=24$)~\cite{PLB_510_17_2001,PLB_622_29_2005}. On the other hand, the $N=34$ sub-shell closure has been suggested in $^{54}$Ca~\cite{Nature_502_207_2013, PRL_121_022506_2018, PRL_123_142501_2019} and in $^{52}$Ar~\cite{PRL_122_072502_2019}. However, it has not been detected in the Sc~\cite{PRC_96_064310_2017}, Ti~\cite{PRL_92_072502_2004,PRC_70_064303_2004, PRC_70_064304_2004} and the Cr~\cite{PLB_510_17_2001,PLB_622_29_2005} isotopes. Parallel to these experimental studies a large amount of theoretical work has been devoted to the analysis of the new magic numbers, both from \textit{ab initio} and phenomenological approaches~\cite{JPG_39_085111_2012, PS_2013_014007_2013, PRC_90_041302_2014,
PRC_90_024312_2014, NPA_694_157_2001, PRC_65_061301_2002}. In particular, most of shell model calculations predict for the titanium isotopes a sub-shell closure in $N=32$ and, depending on the interaction and the many-body method, also in $N=34$. Interestingly axially symmetric beyond mean field theories (BMFT)~\cite{PRL_99_062501_2007} predict the $N=32$ as a closure but not the $N=34$.

Though the mentioned criteria are obviously necessary to consider the sub-shell closures as magic sometimes they are not conclusive and complementary analysis is required. A different way to check new sub-shell closures has been done in Ref.~\cite{PLB_546_55_2002} for the titanium isotopes where the degree of robustness of the $N=32$, 34 shell closures was analyzed based on the experimental Yrast energies of those nuclei. From the naive shell model perspective, the nucleus $^{50}_{22}$Ti$_{28}$ is magic in neutrons and has two protons in the $f_{7/2}$ orbital outside the $Z=20$ core. The valence protons can couple up to angular momentum $I=6 \hbar$,  the $0^{+}_1, 2^{+}_1, 4^{+}_1$ and $6^{+}_1$ states exhibiting an energy pattern typical for seniority isomers. An additional increase of the angular momentum requires either the breaking of the neutron or the proton cores or both. The $8^{+}_1$ state therefore is expected at an excitation energy several MeV above the $6^{+}_1$ state.
In $^{52}$Ti$_{30}$, with two additional neutrons in the $p_{3/2}$ sub-shell, the $8^{+}_1$ can be generated without any core breaking and therefore is expected at lower energy. The nucleus $^{54}$Ti$_{32}$ with four neutrons closing the  $p_{3/2}$ sub-shell could behave similar to $^{50}$Ti$_{28}$ since to reach the $8^+_{1}$ state one should break the $(2p_{3/2})^4$ ($N=32$) sub-shell closure. Similar arguments could also apply to $^{56}$Ti$_{34}$ with the filling of the $p_{1/2}$ sub-shell if the $N=34$ is a closed core as in  $^{54}$Ca. The position and character of the $8^{+}_1$ is therefore very well suited to investigate sub-shell closures in these nuclei.

The purpose of this Letter is twofold, first to perform a theoretical study of the high spin states of the titanium isotopes using state of the art beyond mean field theory (BMFT) recently developed~\cite{PRL_116_052502_2016}. A mean field based study of this problem provides an intrinsic system viewpoint and is complementary to a shell model study. Furthermore, at variance with shell model calculations, the present method does not assume a core and the underlying interaction, i.e., the finite range density dependent Gogny force~\cite{NPA_428_23_1984} is not specifically adjusted to any particular region of the chart of nuclides. 
Second, this study provides a very stringent test to our theory because of three reasons: a) the theoretical description of an aligning pair of nucleons in a single $j$-shell in a BMFT is a very challenging problem because it is a genuine single particle effect, b) the consideration of high angular momentum and c) the simultaneous description of four isotopes.

Modern BMFT's~\cite{RMP_75_121_2003,PPNP_66_519_2011,PS_91_073003_2016,JPG_46_013001_2019} with effective interactions include the restoration of the symmetries broken in the Hartree-Fock-Bogoliubov (HFB) approach~\cite{NPA_383_189_1982,NPA_388_19_1982}, normally angular momentum and particle number conservation, as well as fluctuations around the mean field shape. These are usually considered within the generator coordinate method (GCM), taking as coordinates the quadrupole deformations $(\beta,\gamma)$. This theory known as symmetry conserving configuration mixing (SCCM) have been developed in the last years ~\cite{PRC_78_024309_2008,PRC_81_044311_2010,PRC_81_064323_2010}. These methods based on energy density functionals -Skyrme, Gogny and covariant density functionals- provide, in general, {\em only} qualitative agreement with the experimental spectra. The reason is a stretching of the whole spectrum~\cite{PRC_75_044305_2007,PRC_91_044315_2015}. 
This is related with the lack of an angular momentum dependence in the variational equations to determine the HFB wave functions (w.f.s), see below, which favors  $I=0~\hbar$ states and disfavors the $I\ne0~\hbar$ ones (the larger $I$ the more). Recently we generalized the SCCM methods by including the cranking frequency $\hbar \omega$ as a GCM coordinate for even-even nuclei \cite{PLB_746_341_2015}. Finally in Ref.~\cite{PRL_116_052502_2016}  we extended the range of triaxial quadrupole deformations in the GCM to $-60^{\circ}\leq \gamma \leq 120^{\circ}$, see Fig.~\ref{wav_funct}(e). These improvements not only largely solve the problems of the current BMF approaches but also include single-particle effects through the pair alignment by the cranking procedure making our theory very well suited for the present problem.  
These approaches have been also developed for  odd-even nuclei \cite{EPJA_52_277_2016,PLB_764_328_2017,PRC_98_044317_2018}.


In the SCCM approach the nuclear wave function (w.f.) takes the form
\begin{eqnarray}
|\Phi^{I\sigma}_{M}  \rangle  &=& \sum_{\lbrace\beta,\gamma;\omega;K\rbrace}  f^{I\sigma}_{\lbrace\beta,\gamma;\omega;K\rbrace}
P^{Z}P^{N}P^{I}_{MK}|\phi(\beta,\gamma,\omega)\rangle \nonumber \\
 &=&  \sum_{\lbrace\xi\rbrace}  f^{I\sigma}_{\lbrace\xi\rbrace}|IM;NZ;\lbrace\xi\rbrace \rangle
\label{GCM_ansatz}
\end{eqnarray}
where  $|\phi(\beta,\gamma,\omega)\rangle$ are intrinsic w.f.s discussed below. We have introduced the shorthand notation $\lbrace\xi\rbrace$ and $|IM;NZ;\lbrace\xi\rbrace\rangle$,  with $K$ the projection of the angular momentum on the intrinsic $z$ axis and $(\beta, \gamma)$ and $\hbar \omega$ the generator coordinates for the triaxial shapes and the cranking frequency, respectively. Furthermore  $P^{Z}, P^{N}$ and $P^{I}_{MK}$ are projector operators associated with the particle number and the angular momentum, respectively, see \cite{PRC_81_064323_2010}. The states $|IM;NZ;\lbrace\xi\rbrace \rangle$ are eigenstates of the symmetry operators $\lbrace \hat{I}^{2}, \hat{I}_{z}, \hat{I}_{3}\rbrace$, $\hat{N}$ and $\hat{Z}$. Additionally, $\sigma=1,2,...$ labels the states for a given value of the angular momentum $I$. Hereafter we suppress the labels $N,Z$ to shorten the notation. The coefficients $f^{I\sigma}_{\lbrace\xi\rbrace}$ of the linear combination are found by a minimization of the energy in the Hilbert space spanned by the {\em linearly dependent} w.f.s $|IM;\lbrace\xi\rbrace\rangle$. This variation leads to the Hill-Wheeler-Griffin (HWG) equation
\begin{equation}
\sum_{\lbrace\xi\rbrace}\left(\mathcal{H}_{\lbrace\xi\rbrace,\lbrace\xi'\rbrace}^{I}-E^{I\sigma}\mathcal{N}_{\lbrace\xi\rbrace,\lbrace\xi'\rbrace}^{I}\right)f^{I\sigma}_{\lbrace\xi'\rbrace}=0.
\label{HWG_eq}
\end{equation}
Here we have introduced the norm overlaps $\mathcal{N}^{I}_{\lbrace\xi\rbrace,\lbrace\xi'\rbrace} = \langle IM;\lbrace\xi\rbrace|IM;\lbrace\xi'\rbrace\rangle$ and the Hamiltonian overlap defined by a similar expression. Eq.~(\ref{HWG_eq}) is solved by a two step procedure~\cite{RS.80,PRC_78_024309_2008,PRC_81_064323_2010}: First, the norm matrix is diagonalized, its eigenvalues $n^{I}_k$ and eigenvectors $u^{I}_k(\lbrace\xi\rbrace)$ provide the natural states basis. In the second step the Hamiltonian is diagonalized in this basis providing the eigenvalues $E^{I\sigma}$ of Eq.~(\ref{HWG_eq}) and the eigenvectors $g_{k}^{I\sigma}$. The collective w.f.s $p^{I\sigma}(\beta,\gamma,\omega)=\sum_{k,K} g_{k}^{I\sigma} u^{I}_k(\lbrace\xi\rbrace)$ are orthogonal and consequently the quantities $|p^{I\sigma} (\beta,\gamma,\omega)|^2$ can be interpreted as a probability amplitude. In the $(\beta,\gamma)$ plane the probability amplitude is defined by 
\begin{equation}
|{\cal P}^{I\sigma}(\beta,\gamma)|^2= \sum_\omega |p^{I\sigma} (\beta,\gamma,\omega)|^2.
\label{coll_wf}
\end{equation}
The relevant and only input containing the physics of the problem (apart from the nuclear interaction itself) is the set of HFB w.f.s $|\phi(\beta,\gamma,\omega)\rangle$ of Eq.~(\ref{GCM_ansatz}). These are intrinsic mean field w.f.s that should be determined in the best possible way because they are the building blocks of our theory. In the spirit of the self-consistent HFB method, we should determine different intrinsic w.f.s for each angular momentum. This will lead us to solve at each $(\beta,\gamma)$ point the symmetry conserving HFB equations. A set of integro-differential equations have to be solved iteratively with particle number and triaxial angular momentum projection. This is nowadays a colossal numerical problem with effective interactions like Gogny. The closest feasible approach is to solve the symmetry conserving HFB equations with particle number projection and for different cranking frequencies, instead of performing the exact angular momentum projection before the variation. This method allows the inclusion of intrinsic w.f. with various alignments in the variational space of Eq.~(\ref{GCM_ansatz}). Obviously, the exact AMP is thus performed after the variation, see Eq.~(\ref{GCM_ansatz}).
These premises lead us to minimize the constrained functional
\begin{equation}
 E[\phi]   =   \frac{ \langle \phi | HP^ZP^N|\phi \rangle } {\langle \phi | P^ZP^N|\phi \rangle }
 -  \langle \phi | \omega {\hat J}_x+ \lambda_{q_0} {\hat Q}_{20}+\lambda_{q_2 }{\hat Q}_{22}|\phi \rangle,  
   \label{cranking_1}
\end{equation}
to generate the set of $|\phi(\beta,\gamma,\omega)\rangle$ w.f.s. $\hat{Q}_{2\mu}$ and $\hat{J}_{x}$ are quadrupole moments and the $x$-component of the angular momentum operators, respectively. $\lambda_{q_0}$ and $\lambda_{q_2}$ are the Lagrange multipliers determined by the constraints $\langle \phi |  {\hat Q}_{20}|  \phi \rangle = q_{20}$ and $\langle \phi |  {\hat Q}_{22} |  \phi \rangle =  q_{22}$, while $\omega$ is kept constant during the minimization process. $(\beta,\gamma)$ are defined by $\beta= \sqrt{20\pi(q^{2}_{20} + 2 q^{2}_{22})}/3r^{2}_{0}A^{5/3}$ and $\gamma= \arctan(\sqrt{2}q_{22}/q_{20})$ with $r_{0}=1.2$ fm and the mass number $A$. Through the constraints on $\hat{Q}_{20}$, $\hat{Q}_{22}$ and $\hat{J}_{x}$,  the wave functions $|\phi(\beta,\gamma,\omega)\rangle$, depending parametrically on $(\beta,\gamma,\omega)$, are generated. Specifically, we take three values of the angular frequency, namely, $\hbar \omega =0.0, 0.5$ and $1.0$ MeV, large enough to reach angular momenta larger than $10 \hbar$. For  $\hbar\omega\neq0$ we take 70 points in the $(\beta,\gamma)$ plane, defined by $0\le \beta \le 0.7$ and $-60^{\circ}\le \gamma \le 120^{\circ}$, see Fig.~\ref{wav_funct}(e). We have to consider this larger $\gamma$ interval instead of the usual $0^{\circ}\le \gamma\le 60^{\circ}$ because, due to the term $-\omega {\hat J}_x$ in Eq.~(\ref{cranking_1}), the latter is not enough to describe all rotating shapes, see Fig.~\ref{wav_funct}(e) and Ref.~\cite{PLB_746_341_2015}.  For  $\omega =0$ the three sextants depicted in Fig.~3(e) are equivalent and the intrinsic HFB wave functions $\phi(\beta, \gamma, \omega=0)$ are redundant, i.e., only one sextant needs to be considered for the solution of the HWG equation, Eq.~(\ref{HWG_eq}). Consequently the $(\beta,\gamma)$ probability amplitude of Eq.(\ref{coll_wf}) is given by
\begin{equation}
|{\cal P}^{I\sigma}(\beta,\gamma)|^2= \sum_{\omega \neq 0} |p^{I\sigma} (\beta,\gamma,\omega)|^2 + \frac{1}{3}|p^{I\sigma} (\beta,\tilde{\gamma},\omega=0)|^2
\end{equation}
where $\tilde{\gamma}=|\gamma|$ for $-60^{\circ}\leq \gamma \leq 60^{\circ}$ and $\tilde{\gamma}=120^{\circ} -\gamma$ for $60^{\circ} < \gamma \leq 120^{\circ}$.  We notice that rotations close to $\gamma= -60^{\circ}$ and $\gamma= 120^{\circ}$ are non-collective and can excite single particle degrees of freedom. In the numerical applications the finite range density-dependent Gogny interaction with the D1S parametrization \cite{NPA_428_23_1984} is used together with a configuration space of eight harmonic oscillator shells. Interestingly the incorporation of $\omega$ in the GCM Ansatz of Eq.~(\ref{GCM_ansatz}) is a generalization of the double projection method of Peierls and Thouless~\cite{NP_38_154_1962, PRC_27_453_1983} for the case of rotations. This method is known to provide the exact translational mass in the case of translations. The solution of Eq.~(\ref{HWG_eq}) provides the energies and the wave functions of the  eigenstates of a given nucleus for different values of the angular momentum.

\begin{figure}[t]
	{\centering
		{\includegraphics[angle=-0,width=0.9\columnwidth]{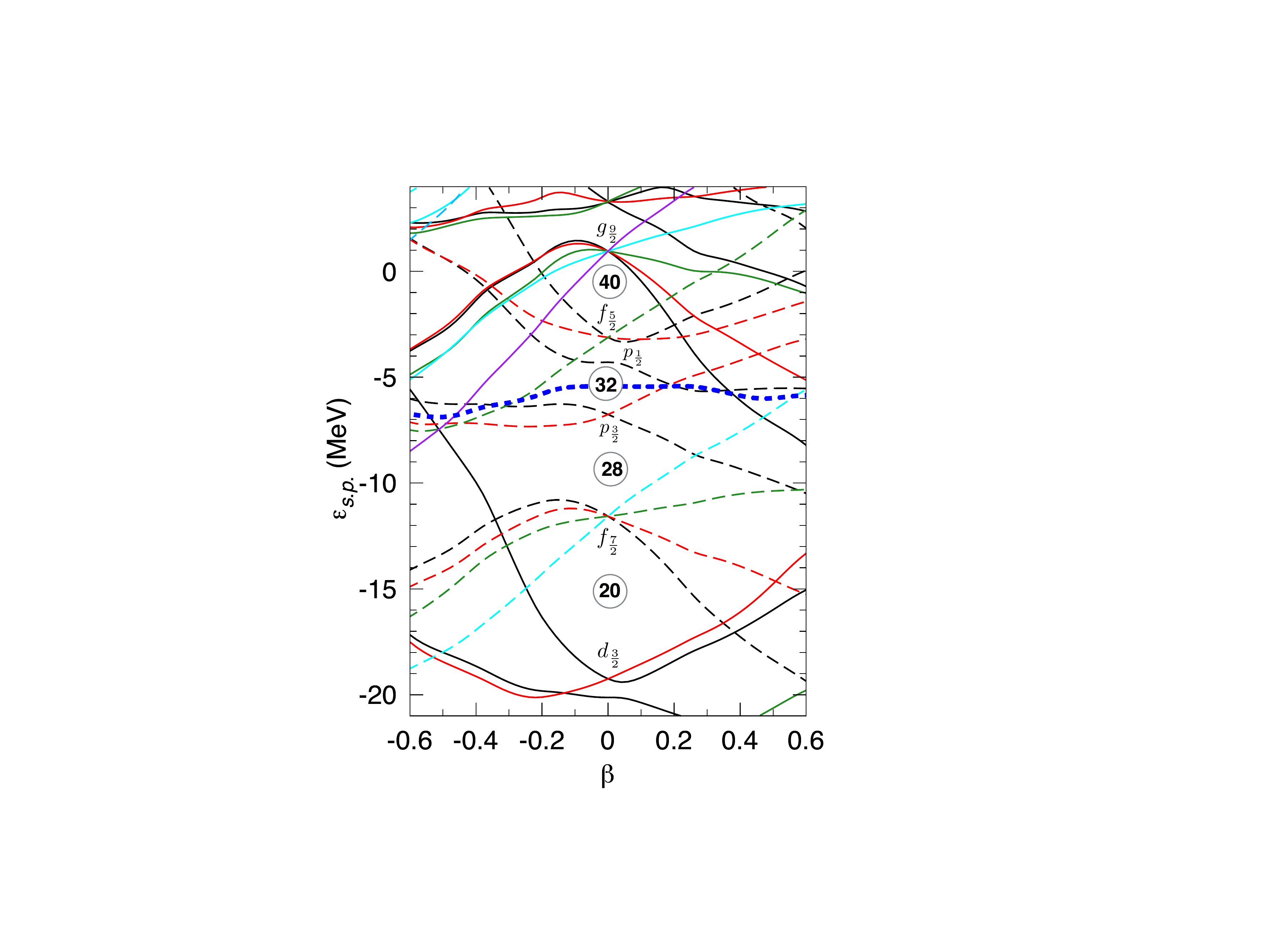}} \par}
	\caption{Neutron single-particle energies of  $^{54}_{22}$Ti along the axial symmetry axis. Solid (dashed) lines correspond to positive (negative) parity levels. Different colors are used to distinguish the third component of the angular momentum quantum number. The thick blue dashed line represents the chemical potential.} \label{54Ti-spe}
\end{figure}

\begin{figure*}[t]
	{\centering
		{\includegraphics[angle=-0,width=1.5\columnwidth]{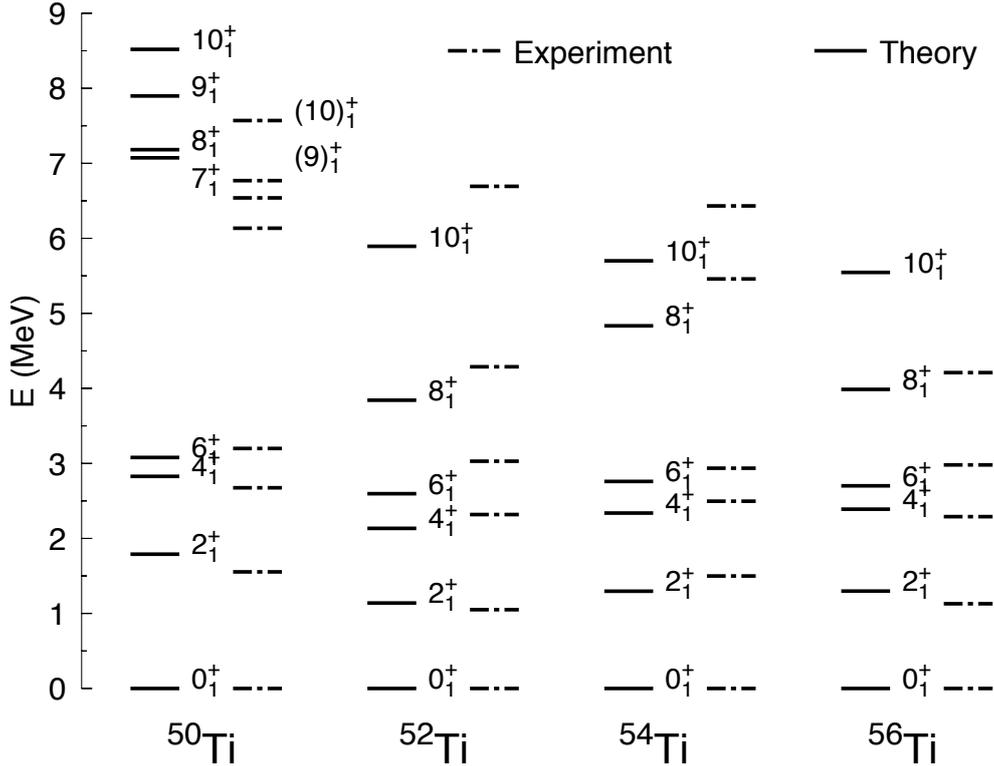}} \par}	
	\caption{Spectra of $^{50-56}$Ti. The theory values are displayed as solid lines and the experimental data~\cite{database} as  dashed lines. } \label{spectra}
\end{figure*}
Let us first discuss some bulk properties of the titanium isotopes under study. The theoretical (experimental~\cite{PRL_120_062503_2018}) binding energy per particle, in MeV, are: 8.81 (8.76) for $^{50}$Ti, 8.74 (8.70) for $^{52}$Ti, 8.62 (8.60) for $^{54}$Ti and 8.48 (8.47) for $^{56}$Ti. The theoretical values show a slight over-binding because the force was fitted globally at the mean field level and the beyond-mean-field correlations provides additional  binding energy~\cite{PRC_91_044315_2015}.  In Table~\ref{Table1} we present the electromagnetic properties of the Yrast states. Unfortunately the experimental data are still scarce. In Ref.~\cite{PRC_62_031301_2000,PLB_633_219_2006} the magnetic moments for  the $2^{+}_1$ states for $^{50-52}$Ti have been studied. The measured values  $\mu/\mu_{N}=  2.89{\it (15)} $ ($^{50}$Ti), and for  $^{52}$Ti, $\mu/\mu_{N}=  1.7{\it (4)} $ are in good agreement with our values. The experimental values in $^{50}$Ti for the electric quadrupole moment   $Q_{exp.}=8 {\it (16)}$ $e$fm$^2$~\cite{NPA_250_381_1975} and the mean squared charge radius $\langle r^2 \rangle^{1/2}=3.57$ fm~\cite{ADNDT_99_69_2013} compare very well with the theoretical values of 5.7 $e$fm$^2$ and 3.6 fm respectively. Notice that no effective charges are used in the calculations.

\begin{table}
	\centering
	\begin{tabular}{|c|c |c| c| c| }
		\hline
		& $^{50}$Ti    &  $^{52}$Ti    &  $^{54}$Ti     &  $^{56}$Ti    \tabularnewline \hline
		$ I^{\pi}_{1}$ 	& $\mu$  $\,\;\;$ Q$_{sp.}$    &  $\mu$  $\,\;\;$ Q$_{sp.}$    &  $\mu$  $\,\;\;$ Q$_{sp.}$    &   $\mu$  $\,\;\;$ Q$_{sp.}$    \tabularnewline \hline		 
		$2^{+}_{1}$       & $2.4  \;\;\;  5.7$  & $2.1 \;\;\;   -10.1$ &  $1.9 \;\;\; 	4.1$ & $1.9  \;\;\;   2.8$   \tabularnewline
		$4^{+}_{1}$       & $5.4 \;\;\;   -3.5$  &  $5.0  \;\;\;  -16.0$ &  $4.6  \;\;\;  -2.5$& $4.6 \;\;\;   -5.4$		\tabularnewline
		${6}^{+}_{1}$     & $8.4  \;\;\;  -26.1$ & $6.9 \;\;\;  -25.7$ & $8.0  \;\;\;  -22.9$ &     $7.6 \;\;\;  -25.1$ 	\tabularnewline
		$8^{+}_{1}$       & $6.4  \;\;\;  -5.7$  & $7.7 \;\;\;  -29.5$ & $7.7 \;\;\;  -32.0$ & $8.8  \;\;\; -27.3$ \tabularnewline
		$10^{+}_{1}$     & $6.6  \;\;\;  -14.6$  & $8.1 \;\;\;  -37.5$ &$8,2  \;\;\;  -31.6$&  $9.6 \;\;\; -29.3$   \tabularnewline
		\hline  
	\end{tabular}
	\caption{Magnetic and spectroscopic electric quadrupole  moments  in units of $\mu_N$ and efm$^2$ respectively,  for the titanium isotopes and several angular momenta.}
	\label{Table1}
\end{table}

 Another important source of information are the single particle-energies (spe), well known in the form of Nilsson plots. To calculate these energies  we have performed axially symmetric calculations with the D1S Gogny interaction in the HFB approach.  Since the spe vary slowly with the mass number it is sufficient  to show them only for one nucleus  
and since we are interested in neutron shell closures we only show the neutron spe   for $^{54}_{22}$Ti  in Fig.~\ref{54Ti-spe}. This plot is also illustrative for the proton spe  since in the Ti isotopes the proton and neutron numbers do not differ much. In this figure we clearly observe large
energy gaps for neutron numbers 20, 28 and 40, a smaller one at 32 and a little one at 34. These are mean field predictions for the candidates for shell closures.  One knows, however, that the coupling to additional degrees of freedom may provide energy shifts that modify this picture.  Obviously, such modifications will mainly affect the small energy gaps.

Let us now discuss the solution of Eq.~(\ref{HWG_eq}). 
The energies of the lowest excited states of the titanium isotopes are plotted in Fig.~\ref{spectra}, together with the experimental values~\cite{database}. Taken into account that no parameter fit has been done for this calculation{\tiny }, the agreement between theory and experiment is very good. The energies of the $2^{+}_{1}$ states suggest $^{54}$Ti$_{32}$  as a sub-shell closure candidate and $^{56}$Ti$_{34}$  as a potential one. The spectrum of the nucleus $^{50}_{22}$Ti$_{28}$ displays a very characteristic feature, namely, compressed as a function of $I$ for  $I\leq 6 \hbar$ and with a large energy gap between the   $I\leq 6 \hbar$ and  $ I \geq 7 \hbar$  states. The interpretation is simple, see Fig.~\ref{54Ti-spe}, the two protons in the  $1f_{7/2}$ shell outside the $Z=20$ core can maximal couple to $I=6 \hbar$ without breaking the proton and/or neutron cores. This behavior can be found in many nuclei with two valence particles (holes) and an inert core. For example, in $^{42}$Ca,   $^{210}$Pb,  $^{210}$Po~\cite{database}, and even in exotic nuclei  like  $^{130}$Cd \cite{PRL_99_132501_2007}. This characteristic, however, is not expected in the spectrum of $^{52}$Ti$_{30}$ with four valence particles in open shells, because additionally to the two protons we now have two neutrons in the  $2p_{3/2}$ shell outside the $N=28$ shell closure, see Fig.~\ref{54Ti-spe}. A glance at Fig.~\ref{spectra} confirms this assumption, the spectrum is not much compressed as a function of $I$, in particular the energy of the $2^{+}_{1}$ state is considerably lower than in $^{50}$Ti and the $6^{+}_{1}-8^{+}_{1}$ energy gap is drastically reduced. The $(2p_{3/2})^4$ configuration of $^{54}$Ti$_{32}$, see Fig.~\ref{54Ti-spe}, is reflected in the spectrum of this nucleus: The energy of the $2^{+}_{1}$ state increases and so does the $6^{+}_{1}-8^{+}_{1}$ energy gap.  Based on the increase of the $2^+_{1}$ energy the $N=32$ number has been proposed as a new magic number~\cite{PLB_546_55_2002,PRC_70_064303_2004}. We can see in Fig.~\ref{spectra} that the results at large values of the angular momentum seem to support this thesis. It remains to analyze the theoretical interpretation of the $8^+_{1}$ state to confirm it. Lastly, in the pure single particle shell model, the nucleus $^{56}$Ti$_{34}$ is obtained by adding two neutrons in the $2p_{1/2}$ sub-shell to the nucleus $^{54}$Ti that close this orbit, see Fig.~\ref{54Ti-spe}. This could lead to the interpretation of $N=34$ as another new magic number, taking into account that the energy of the $2^+_{1}$ is also higher than in $^{52}$Ti. However, in this case the $6^{+}_{1}-8^{+}_{1}$ gap is reduced similarly to the $^{52}$Ti case. Again the examination of additional properties of the $8^{+}_{1}$ state would clarify this question.
\begin{figure*}[t]
{\centering
{\includegraphics[angle=0,width=1.8\columnwidth]{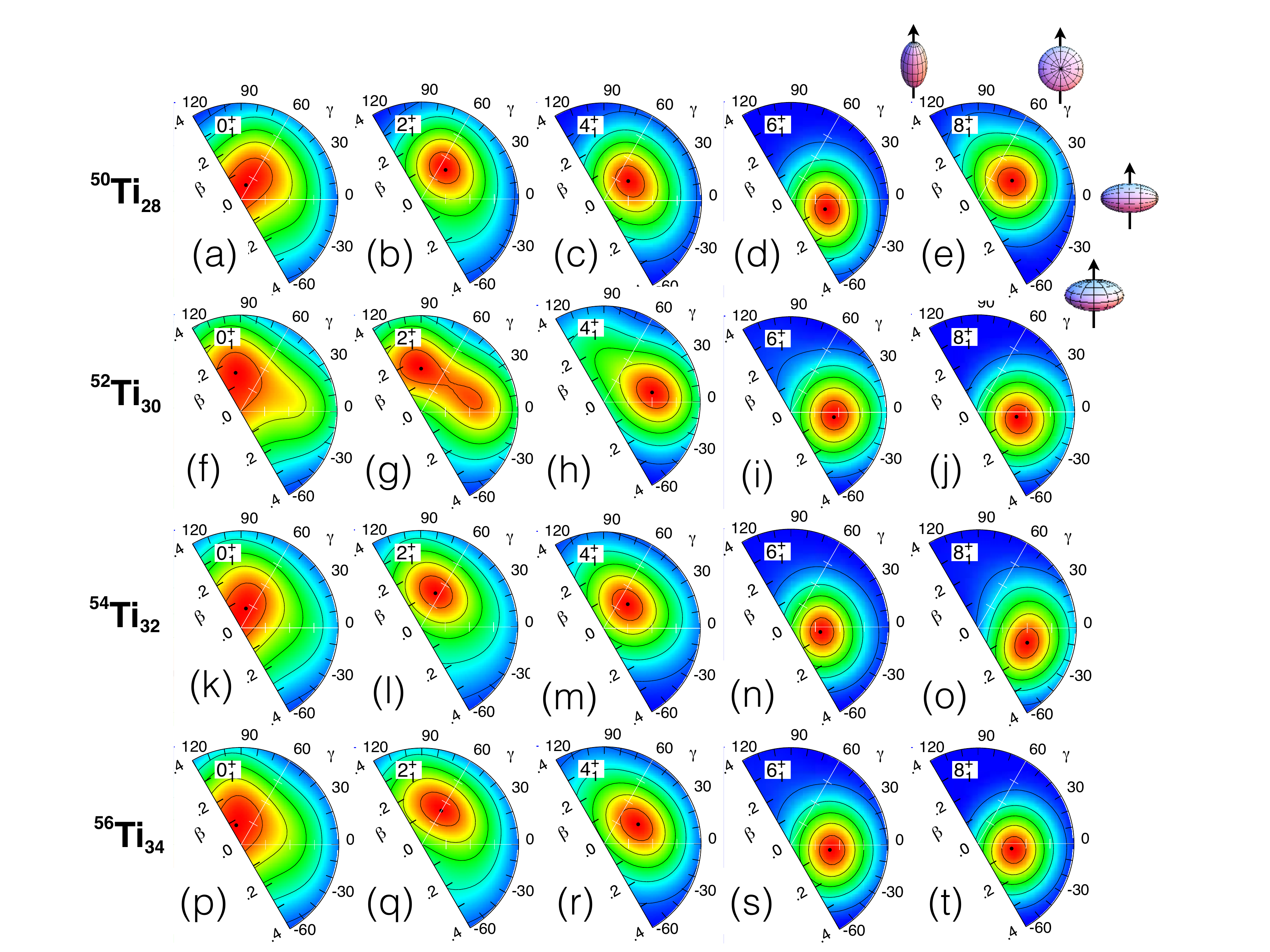}} 	\par}
\caption{(Color online) Squared collective wave functions in the  $(\beta,\gamma)$ plane for several states of the titanium isotopes. In each plot the value of the outer contour corresponds to one-fifth of the maximum value plotted as a bullet. Each contour is incremented by this amount up to the maximum value. Panel e) illustrate the shapes for selected $(\beta, \gamma)$ values as well as the rotation axis. The angle $\gamma$ is given in degrees. }
\label{wav_funct}
\end{figure*}

Crucial information about the structure of the states is provided by the collective wave w.f.s, Eq.~(\ref{coll_wf}). In Fig.~\ref{wav_funct} we show the evolution of the wave functions of the Ti isotopes with the angular momentum. For the sake of clarity we only show the relevant part, although our calculations extend up to $\beta=0.7$ values. The w.f. of the $0^+_{1}$ states, panels (a)-(f)-(k)-(p), have a broad and flat maxima mostly extending from the oblate shape ($60^{\circ}$) to the non-collective rotation axis ($120^{\circ}$), see panel  Fig.~\ref{wav_funct}(e). This is the non-collective sextant. The nucleus $^{52}$Ti shows, however, some prolate components and $^{56}$Ti an incipient collectivity. We would like to mention that the $\omega$ mixing in Eq.~(\ref{coll_wf}), is always large. For example, in the $0^+_{1}$ states is around $40\%$ thus causing single particle alignments. 

There are, in general, two main ways to generate additional angular momentum. If the nucleus keeps a spherical shape, one has to promote particles to higher shells with larger angular momentum. In the Ti isotopes the closest shells to the Fermi surface do have small angular momentum and this procedure will not be very efficient. The second mode is to deform the nucleus because the deformation favors the lowering of high angular momentum orbitals mixing them with those single-particle orbits close to the Fermi level. Increasing the angular velocity $\omega$ allows for both mechanisms through the alignment of particles by non-collective rotation and by the centrifugal force inducing collective deformation. 

The w.f.s of the $2^+_{1}$ states of the nuclei $^{50}$Ti and $^{54}$Ti, panels (b)-(l), show maxima with a larger $\beta$-deformation, around $0.15$, and close to the oblate axis. The nucleus $^{52}$Ti, panel (g), on the other hand, display strong triaxial  components also in the collective sextant, explaining the low excitation energy of this state. Finally $^{56}$Ti, panel (q), exhibits slightly more collective w.f.s than  $^{50}$Ti and $^{54}$Ti. The spectroscopic moments displayed in Table~\ref{Table1} for  the $2^{+}_{1}$ states are in line with the previous discussion.
 
 The maxima of the w.f.s of the $4^+_{1}$ states move to collective shapes following a comparable pattern as the $2^+_{1}$ ones: $^{50}$Ti and $^{54}$Ti look very similar, $^{52}$Ti very collective and, in between, the $^{56}$Ti isotope.  As we can see in Table~\ref{Table1} the quadrupole moment of the $4^{+}_{1}$ states turn prolate. The w.f.s of the $6^{+}_{1}$ states look all four very similar and the quadrupole moments reach values around $-25 $ $e$fm$^2$. The fact that all $6^{+}_{1}$ states do have approximately the same quadrupole moment is an indication that the angular momentum is made mostly by the two valence protons, the centrifugal force driving the nucleus to this aligned configuration in all four nuclei. This is supported by the expectation values of $\hat{I}^{2}$ with the wave functions of Eq.~(\ref{GCM_ansatz}), shown in Table~\ref{Table2} separately for protons and neutrons and the fact that their w.f.s shown in panels (d), (i), (n) and (s) look quite similar. The maximum of the w.f. of the $6^{+}_{1}$ states of $^{50-56}$Ti being located at $(\beta,\gamma)$ values $(0.15, -16^\circ), (0.16,-7^{\circ}),(0.12, -9^\circ)$ and $ (0.16,-7^{\circ})$ respectively.  
The interesting question now is how the different nuclei build in the additional spin needed for the state $8^{+}_{1}$. In panels Fig.~\ref{wav_funct}(e), (j), (o) and (t) we show the associated w.f.s. For $^{50}$Ti the w.f. maximum of this state appears at $(0.16, 30^\circ)$, i.e., a radical change to a triaxial nucleus with a very small value of the quadrupole moment, see Table~\ref{Table1}. In $^{52}$Ti, the maximum of the w.f. is located at $(0.15, -8^{\circ})$, i.e., quite close to the maximum of its $6^{+}_{1}$ state, indicating that no structural change is required to generate additional angular momentum. For $^{54}$Ti the maximum shows up at $(0.21, -17^{\circ})$, i.e., an increase of $75\%$ in the deformation of the nucleus as compared to the $6^{+}_{1}$ state. Hence, a noticeable structural change is observed from the $6^{+}_{1}$ to $8^{+}_{1}$ states, similarly to $^{50}$Ti. Lastly, the $8^{+}_{1}$ w.f. of $^{56}$Ti presents the maximum at $(0.14, -8^{\circ})$, again quite close to the $6^{+}_{1}$ case, i.e., no major changes are needed to generate angular momentum. These results corroborate that $N=32$ is a sub-shell closure and $N=34$ does not. 
 In terms of Fig.~\ref{54Ti-spe} we can conclude that the coupling of the spe to other degrees of freedom does not modify considerably the results of the HFB concerning shell closures and second that a relatively small gap of the size  of $N=32$ in  Fig.~\ref{54Ti-spe} is able to provide characteristics of a magic number.

\begin{table}
	\centering
	\begin{tabular}{|c|c |c| c| c| }
		\hline
		& $^{50}$Ti    &  $^{52}$Ti    &  $^{54}$Ti     &  $^{56}$Ti    \tabularnewline \hline
		$ I^{\pi}_{1}$ 	& Z  $\,\,\,\,\,$ N    &  Z  $\,\,\,\,\,$ N  &  Z  $\,\,\,\,\,$ N   &  Z $\,\,\,\,\,$  N   \tabularnewline \hline		 
		$2^{+}_{1}$       & $4.5  \;\;\;  1.5$  & $3.4 \;\;\;   2.6$ &  $3.7 \;\;\; 	2.3$ & $3.4  \;\;\;   2.6$   \tabularnewline
		$4^{+}_{1}$       & $17.1 \;\;\;    2.9$  &  $14.0  \;\;\;  6.0$ &  $14.2  \;\;\;   5.8$& $14.1 \;\;\;   5.9$		\tabularnewline
		${6}^{+}_{1}$     & $39.5  \;\;\;   2.5$ & $32.7  \;\;\;  9.3$ & $37.3   \;\;\;  4.7$ &     $34.5 \;\;\;  7.5$ 	\tabularnewline
		$8^{+}_{1}$       & $42.9  \;\;\;  29.1$  & $49.0 \;\;\;  23.0$ & $48.0  \;\;\;  24.0$ & $49.4  \;\;\; 22.6$ \tabularnewline
		$10^{+}_{1}$     & $61.3  \;\;\;  48.7$  & $65.2 \;\;\;  44.8$ &$65.1  \;\;\;  44.9$&  $65.0  \;\;\; 45.0$   \tabularnewline
		\hline  
	\end{tabular}
	\caption{Proton (Z) and neutron (N) contribution to the expectation value of $\langle \Phi^{I}_{M}|\hat{I}^2|\Phi^{I}_{M}  \rangle$ in units of $\hbar^2$, i.e., $ I(I+1) \hbar^2$,  for the titanium isotopes and several angular momenta.}
	\label{Table2}
\end{table}

\begin{figure}[t]
{\centering
{\includegraphics[angle=0,width=1.\columnwidth]{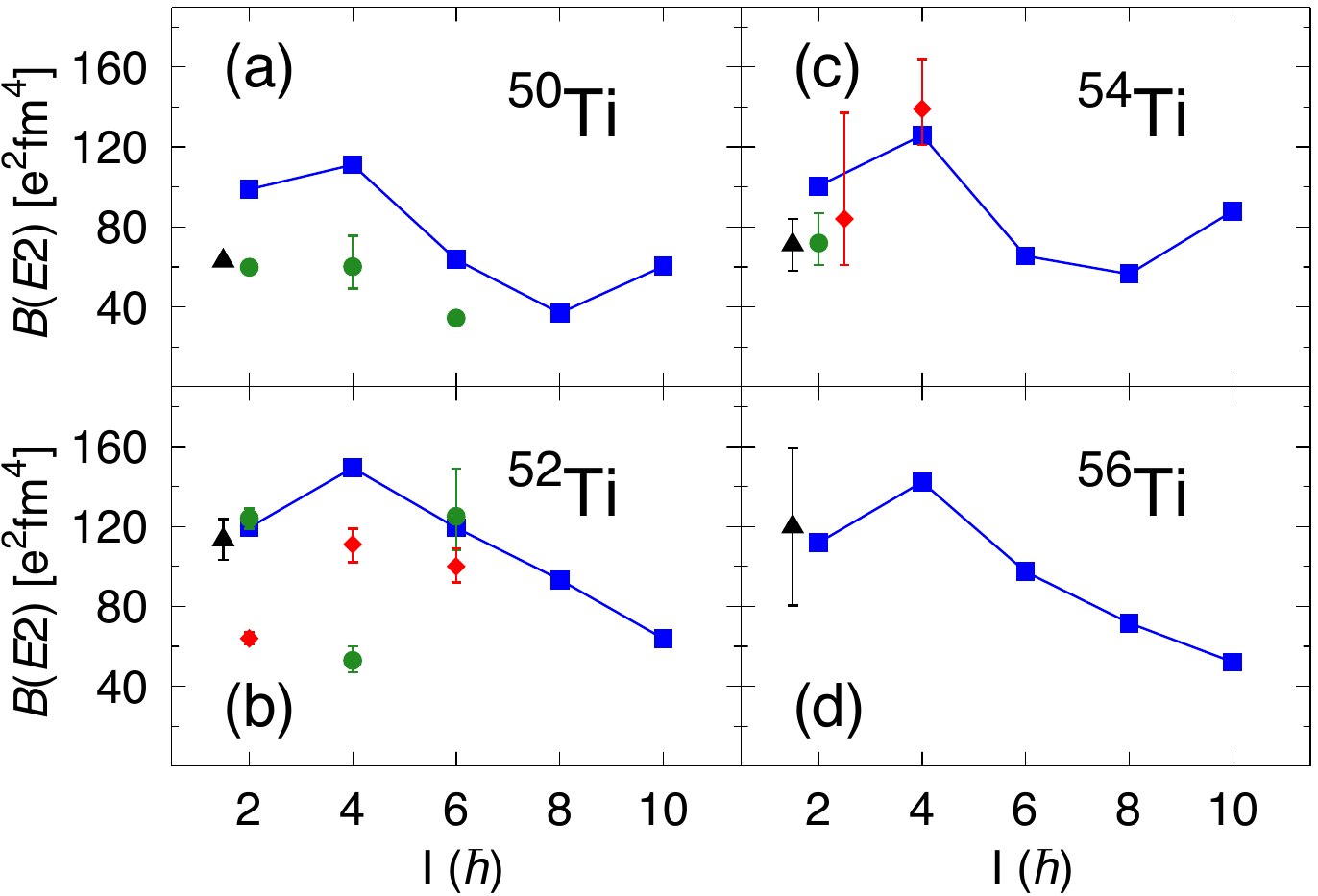}}
\par}
\caption{(Color online) The reduced  transition probabilities $B(E2; I\longrightarrow I-2)$ for $^{52-54-56}$Ti as a function of the angular momentum. The experimental values have been taken from Ref.\cite{PRC_71_041302_2005} (black triangles), Ref.~\cite{PLB_633_219_2006} (green bullets) and Ref.~\cite{Goldkuhle} (red diamonds).}
\label{trans}
\end{figure}

The second observable relevant to the sub-shell closures are the $B(E2; I \rightarrow I-2)$ reduced transition probabilities. These are plotted in Fig.~\ref{trans} as a function of the angular momentum. The theoretical values are in line with the previous discussion. Thus the transitions probabilities are larger for $^{52}$Ti and $^{56}$Ti than for $^{50}$Ti and $^{54}$Ti, a clear indication of the stronger collectivity of the former ones. The experimental values for the $B(E2)$ transitions probabilities in $^{50}$Ti, see panel  Fig.~\ref{trans}(a), have a similar behavior as in the theory. The theoretical values, however, are larger than the experimental ones. This is a well known characteristic of the BMFT that predict larger collectivity in double shell closed nuclei than experimentally measured. In panel Fig.~\ref{trans}(b) the values for $^{52}$Ti are shown. For the transitions $2^{+}_{1} \rightarrow 0^{+}_{1}$ there are three experimental values, two in very good agreement with the theory. For $6^{+}_{1} \rightarrow 4^{+}_{1}$ there are two experimental values in good agreement with the theory. The experimental values for the  $4^{+}_{1} \rightarrow 2^{+}_{1}$ transition are clearly incompatible with each other and smaller than the theoretical value. The $B(E2)$ values for $^{54}$Ti are plotted in panel Fig.~\ref{trans}(c) and we find good agreement between theory and experiment. Finally, for $^{56}$Ti the only experimental result is in excellent agreement with the theoretical one. Interestingly the bending observed at high spins in $^{50}$Ti and $^{54}$Ti does not show up in $^{52}$Ti nor in $^{56}$Ti. 

In conclusion, in this Letter we analyze the new subshell closures $N=32$ and $N=34$ in the titanium isotopes and their robustness at high angular momentum.  In our study we use a symmetry conserving configuration mixing theory which include triaxial shape fluctuations combined with the double projection method of Peierls and Thouless to properly deal with the angular momentum dependence of the excited states. Our calculations provide a good description of the basic relevant bulk properties like binding energies, magnetic dipole moment, spectroscopic electric quadrupole moments and radii of these nuclei as well as excitation energies and transition probabilities. For the analysis of  the calculations of the $N=30$, 32 and 34 titanium isotopes we take as a reference the behavior at high spin of the $N=28$ shell closed nucleus $^{50}$Ti. We conclude that $N=32$ remains as a magic number at high spin, this being a clear manifestation of the robustness of this magic number. The hypothetical $N=34$ sub-shell closure, however, does not satisfy the same criteria. The advantages of our approach are the added value of the intrinsic system interpretation of the BMFT to pin down the configuration changes and that our interaction, the Gogny force, is well known for its predictive power and good performance for bulk properties all over the chart of nuclides.

This	work	was	supported	by the Ministerio de Econom\'ia y Competitividad under contract  FPA2014-57196-C5-2-P and the Ministerio de Ciencia, Innovaci\'on y Universidades under contract PGC2018-094583-B-I00.

\end{document}